\documentclass[
reprint,
 amsmath,amssymb,
 aps,
floatfix,
]{revtex4-2}
\usepackage{graphicx}
\usepackage{dcolumn}
\usepackage{bm}

\begin{document}

\preprint{APS/123-QED}

\title{Towards Generation of Indistinguishable Coherent States} 
\author{Pranshu Maan\\Robert Bosch LLC}
\email{Pranshu.Maan@us.bosch.com\\15000 N Haggerty road, Plymouth, Michigan-48170, USA}
\date{\today}

\begin{abstract}
We have addressed the characteristic distinguishability of coherent states in the temporal domain from a directly modulated quantum well-based gain-switched laser diode. Using small-signal and large-signal models, we identify tunable parameters to generate indistinguishable coherent states from an electrically pumped semiconductor laser. The experiment confirms the generation of indistinguishable signal and decoy coherent states as predicted by the numerical simulation.
\end{abstract}

\maketitle
\section{Introduction}
Widely used asymmetric and symmetric cryptography is intractable problems based on complexity conjecture P$\neq$NP [3]. Albeit based on conjecture, this cryptography is more about how long it will take to solve the problem using the brute force method using unlimited resources. Due to Shor's algorithm, tools like quantum computers can render public-key cryptography insecure [17]. Moreover, this insecurity cannot be physically detected should the public key get compromised.

Quantum Cryptography provides hardware-based information-theoretic security, providing eavesdropper detection without underlying conjectures. One of the implementations of Quantum Cryptography is Quantum Key Distribution (QKD), which allows the exchange of random keys on the quantum channel. An ideal QKD provides a provable secure implementation. However, an imperfection in these implementations often opens doors to side channels [9][10][12]. Ideally, the source for QKD should emit one photon at a time and on-demand. Unfortunately, single-photon sources are expensive and have low collection efficiency, making them as good as alternative sources. Moreover, a high rate single-photon source is yet to be demonstrated [4]. Implementation of QKD using alternative sources has been widely discussed in [2], [11], [14].

We propose a generation of indistinguishable signal–decoy-state in the temporal domain by using a gain-switched single laser diode for each polarization encoded qubit. The behavior of a free-running laser diode at room temperature for a low-cost QKD framework has been studied. As discussed in [1], this work addresses the challenge of implementing decoy-state-based QKD using direct laser modulation by fine-tuning the shape of the injection. We have theoretically estimated parameters to control distinguishability for small signal excitation and have extended our analysis to large-signal excitation by numerical solving the laser rate equation for the quantum well-based laser diodes. Further, we theoretically characterize the emission of a laser diode and explain its generation from a vacuum state. Trend predicted by theoretical estimation for small signal excitation and numerically simulation for large-signal excitation agree with the experimental results, and indistinguishability of signal and decoy-state are quantified.
\section{Decoy state based Quantum Key Distribution}

A Weak Coherent Source from an attenuated laser diode is a widely used photon source in QKD. In a single electromagnetic mode of the photon, coherent state $|\alpha\rangle$ can be represented as (in the ideal case; practically, it will be a mixed state):

\begin{eqnarray}\label{eq:coherent}
|\alpha \rangle=e^{-{\dfrac{|\alpha|^{2}}{2}}}\sum^{\infty}_{n=0}\dfrac{\alpha^{n}}{\sqrt{n!}}|n\rangle
\end{eqnarray}
where~$\alpha$ is a complex number and~$n$ represents the single mode photon number

The density matrix for the coherent field state in the $|n\rangle$ basis state can be written as:
\begin{eqnarray}\label{eq:density}
\rho=\sum \rho_{nm}|n\rangle \langle m|
\end{eqnarray}
where
\begin{eqnarray}\label{eq:matrix}
\rho_{nm}=\langle n|\alpha \rangle\langle \alpha|m\rangle=\dfrac{\alpha^{n}\alpha^{*m}}{\sqrt{n!m!}}e^{-|\alpha|^{2}}
\end{eqnarray}
Then the probability distribution~$P_{n}$ of an ideal coherent source in the number state can be written as:
\begin{eqnarray}\label{eq:probability}
P_{n}=\rho_{nn}=\dfrac{|\alpha|^{2n}e^{-|\alpha|^{2}}}{n!}
\end{eqnarray}
The theoretical bound for the probability of obtaining more than one photon can then be approximated from Eqn.~\eqref{eq:probability} as a conditional probability :
\begin{eqnarray}\label{eq:multi}
P_{multi-photon}\cong \dfrac{|\alpha|^{2}}{2}
\end{eqnarray}
Where $\alpha$ is the eigenvalue of the destruction operator representing the mean number of the photon in a laser pulse; this is one of the major drawbacks and a security flaw because the practical implementation of the coherent source emits multiple photons. Consequently, if Alice sends a multi-photon state to Bob, the eavesdropper can split the multi-photon state, retain one copy of Alice's state and send the remaining state to Bob while selectively suppressing the signal state without getting detected(Photon Number Splitting attack). Finally, when Alice and Bob exchange basis selection on a public channel, Eve will have access to the entire/partial key.

In order to overcome this drawback, decoy state-based QKD was proposed [15]. In a decoy-state implementation, a pulse whose average photon number is different from that of the signal pulse is sent together at random. These pulses differ only in the mean number of photons but have indistinguishable characteristics like pulse width, frequency, etc. Alice can transmit either a signal or decoy-state to Bob by simply varying the mean number of the photon (intensity) of the WCS pulse, as defined in Eqn.~\eqref{eq:probability}. If $Y_{j}$ represents yield by Bob's detector to a pulse sent by Alice containing~$j$ photons, i.e., the conditional probability of detection of the photon at Bob's receiver when Alice sends a signal with~$j$ number of photons, then:
\begin{equation}\label{eq:yield}
Y_{j}=1-(1-Y_{0})(1-\eta)^{j}
\end{equation}
Where $Y_{0}$ is the yield with zero photon, i.e., representing the dark count rate, and $\eta$ is the transmission probability of signal containing~$j$ photons. Considering the probabilistic distribution of photons in weak coherent pulse, we can define the gain $Q_{j}$ of detecting pulse for Bob with~$j$ photons sent by Alice as:
\begin{equation}\label{eq:gain}
Q_{j}=Y_{j}\dfrac{|\alpha|^{2n}e^{-|\alpha|^{2}}}{n!}
\end{equation}
Then overall gain or probability of a pulse with the mean number of photon~$n$ at Bob's receiver is:
\begin{equation}\label{eq:overall}
Q=\sum_{j}^{n}Y_{j}\dfrac{|\alpha|^{2j}e^{{-|\alpha|^{2}}}}{j!}
\end{equation}
If $\eta$ represents the overall efficiency of the bit transmission from Alice to Bob, including channel efficiency, then the error rate~$e_{j}$ of detecting a pulse with~$j$ photons considering the setup efficiency, dark count error yield of~$e_{darkcount}$ and Bob's detection efficiency~$e_{detector}$, is defined as:
\begin{equation}\label{eq:error}
    e_{j}=\dfrac{e_{darkcount}+e_{detector}\eta}{Y_{j}}
\end{equation}
QBER~$E_{j}$ for each yield is given by:
\begin{equation}\label{eq:QBER}
    E_{j}=e_{j}Y_{j}\dfrac{|\alpha|^{2n}e^{-|\alpha|^{2}}}{n!}
\end{equation}
and overall QBER is expressed as:
\begin{equation}\label{eq:QBER1}
  E Q=\sum_{j}^{n}e_{j}Y_{j}\dfrac{|\alpha|^{2j}e^{-|\alpha|^{2}}}{j!}
\end{equation}
Alice and Bob can experimentally determine the gain~$Q$ and the overall QBER~$EQ$. From Eqn. \ref{eq:overall} and Eqn. \ref{eq:QBER1},~$e_{j}$ and~$Y_{j}$ for a pulse containing~$j$ photons can be estimated and acceptable ranges can be determined. Any attempt by Eve will affect~$e_{j}$ and~$Y_{j}$ beyond the acceptable range, thereby revealing its presence.

Based on the equation for yield and error rate, the secure key rate for WCS based QKD can be written as [22]:
\begin{equation}\label{eq:keyrate}
    S\geq q(-Q_{\mu}f(E_{\mu})H_{2}(E_{\mu})+Q_{1}[1-H_{2}(e_{1})])
\end{equation}
Where~$S$ is the key rate,~$q$ = the ratio of signal state for both Alice and Bob to the total number of pulses sent by Alice. Mean photon number $\mu$ depends on injection current, $Q_{\mu}$is the fractional yield rate, and $E_{\mu}$ represents the error rate of the signal state detected by Bob. $Q_{1}$ and $e_{1}$ are fractional yield and error rates of the decoy detection by Bob for a single-photon state. $H_{2}$ is the Shannon information entropy which represents a statistical fluctuation in error rate due to pulse with $\mu$ photons, and $f(E_{\mu})$ represents the error correction function. $Q_{\mu}, E_{\mu}$ can be obtained as discussed in Eqn. [8]-[12].

In-order to detect presence of Eve, indistinguishability between signal state and decoy state is one the fundamental requirements. Consequently, Eve can only know number of photons in each pulse detected, but should not be able to differentiate between whether the photon detected was from signal state or decoy state. If the yield and the error rates are tightly bounded, it will then be possible to detect presence of Eve because any attempt by Eve to detect these states will impact statistical distribution of signal and decoy states.

For signal state and decoy-state to be completely indistinguishable, they must have the same spatial, spectral and temporal behavior. One of the exciting ways to implement this will be using photons from resonantly excited systems such as quantum dots and 2D materials[16], which results in Fourier transform limited characteristics of the emitted photon.
Implementations of signal-decoy-state have been analyzed by [1], [2]. The decoy-state can be implemented either by using an external intensity modulator with a dedicated driver circuit or by varying injected pump current in the laser diode [1]. In the former case, a separate optical attenuator and a dedicated driving circuit are needed. It impacts the overall form factor of the implementation, and adds to cost. In the latter case, indistinguishability was lost due to relaxation oscillation of the laser diode and delay in emission w.r.t the injected current.
Additionally, due to the distinguishability of pulse generated from the injection current pumping technique, the security rate was zero, i.e., the transmission was insecure. Implementation mentioned in [2] needs two diodes for each basis state. Eight diodes are needed in a specially constructed holder for four basis states. This arrangement adds to cost and restricts the selection of eight laser diodes and the positioning of laser diodes in the setup. In this work, for a smaller form factor and cost of implementation, we propose the generation of the indistinguishable signal and decoy-state using direct modulation of the laser diode.

\section{Theory}
Direct modulation of the injection pumps current results in distinguishability of signal and decoy-state[1]. This distinguishability is due to relaxation oscillation and characteristic delay between the signals. Characterization of this relaxation oscillation has been widely described in detail [8],[11]. An increase in injection current increases photon emission; as a result, the electron carrier density rate decreases. At a certain electron density, the rate of generation of photons starts to decrease and contribution of stimulated emission decreases, increasing carrier concentration. This cycle results in relaxation oscillation until it achieves a steady-state level. These distinguishable attributes in the pump current will have a signature in the laser's output pulse, i.e., if there are multiple peaks in the injection current, laser output will follow this profile. 

\begin{figure}[h]
\includegraphics[width=8cm]{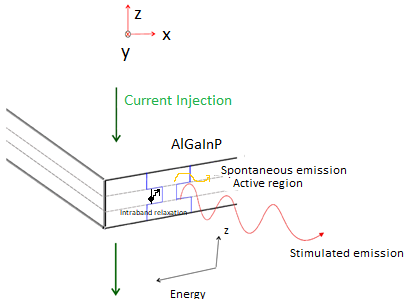}
\caption{\emph{General representation of quantum well-based laser diode: Current injection, spontaneous emission, stimulated emission, and intra-band relaxation change carrier concentration and governs the rate of emission from the laser diode. Above the injection threshold, stimulated emission dominates, the phenomenon of spontaneous emission is dominant below the threshold. Photon density in the active region builds up when emission overcomes the cavity loss.We have considered carrier lifetime~$\tau_{n}$ and photon lifetime~$\tau_{p}$.}}
\label{fig:quantumwelllaserdiode}
\end{figure}

We generate indistinguishable signal state and decoy-state by direct modulation and temporal tuning of injection current into an AlGaInP-based laser diode with heterojunction quantum well, as shown in FIG.~\ref{fig:quantumwelllaserdiode}.

Wave-function for an exciton: electrons in the conduction band and holes in valence band at lattice point~$i$ and~$j$ in a quantum well, with the wave vector $k_{e}$ and $k_{h}$, can be represented as:
\begin{equation}\label{eq:exciton}
    |\psi(t)\rangle=|\psi^{i,j}_{n}\rangle=\sum_{k_{e},k_{h},i,j}\psi_{k_{e},k_{h},i,j}|k_{e},i,k_{h},j\rangle
\end{equation}
And individual wave-function of the electron and hole in the periodic lattice potential at a lattice point~$i$ and~$j$ respectively can be expressed as the Fourier transform:
\begin{eqnarray}\label{eq:EHP}
|\psi_{e,i}\rangle=\dfrac{1}{\sqrt{NV_{c}}}\sum e^{ik.R} \phi_{i}(r)\\
|\psi_{h,i}\rangle=\dfrac{1}{\sqrt{NV_{c}}}\sum e^{ik.R} \phi_{j}(r)
\end{eqnarray}
This represents a Bloch function where~$\phi$ represents variation in local potential around lattice point,~$r$ is the relative position of e-h pair,~$N$ is the number of cells, and~$V_{c}$ is the volume of the cell.
For the exciton wavefunction, the density matrix can then be represented as:
\begin{equation}\label{eq:dmatrix}
    \rho=\sum P|\psi^{i,j}_{n}\rangle \langle \psi^{i,j}_{n}|
\end{equation}
Considering probability distribution~$P$ of electron and hole be time independent, rate of change of density matrix representing change in electron distribution can be represented as:
\begin{equation}\label{eq:edistribution}
\dfrac{d\rho}{dt}=\sum P\dfrac{|\psi_{e,i}\rangle}{dt}\langle \psi_{e,i}|+P|\psi_{e,i}\rangle\dfrac{d\langle\psi_{e,i}|}{dt}
\end{equation}
From time-dependent Schrodinger equation and expressing Hamiltonian~$H_{e}$ for optical wave interacting with the electron in quantum well as:
\begin{equation}\label{eq:hamiltonian}
    H_{e}=\dfrac{1}{2m}(p-qA)^{2}+qU(r)
\end{equation}
where~$A$ is the vector potential associated with the electromagnetic field,~$U(r)$ is the potential energy of the  electron,~$m$ and~$p$ are the mass and momentum of the particle, and for the isolated system, the time evolution of density matrix representing the interaction between optical wave and electron wave-function in a laser cavity can then be derived as:
\begin{equation}\label{eq:dmatrixhamiltonian}
    \dfrac{d\rho}{dt}=\dfrac{1}{i\hbar}[H_{e}-er.E,\rho]
\end{equation}
where~$H_{e}$ is the Hamiltonian of the electron,~$r$ is the position vector. For the derivation of the rate equation,~$E$ is a classical electric field. Eqn.~\ref{eq:dmatrixhamiltonian} represents the interaction between optical wave and electron wave-function in laser cavity resulting in stimulated emission, which further change electron distribution.

Current pumping in laser diode results in population inversion. Additionally, based on the dynamic equation for diagonal and off-diagonal elements, the rate of change of electron density in the conduction band due to light-matter interaction, intra-band relaxation, spontaneous emission, and current pumping can be written as [13]:
\begin{widetext}
\begin{equation}\label{eq:rateeqn1}
   \dfrac{dN_{cb}}{dt}=N_{0}\sum_{cb}\dfrac{i}{h}(R_{nm}\rho_{mn}-\rho_{mn}R_{nm})E-N_{0}\sum\dfrac{\rho_{nn}-\rho_{b}}{\tau_{b}} -N_{0}\sum\dfrac{\rho_{nn}-\rho_{0}}{\tau_{spon}}+N_{0}\sum_{i} \Delta_{i} 
\end{equation}
\end{widetext}
where $N_{cb}$ is the electron distribution in conduction band $\Delta_{i}$ is the charge injection, $N_{0}$ is the total electron density conduction and valence band, $R_{nm}$ is the dipole moment of charge particles, $\rho_{0}$ and $\rho_{b}$ represents electron distribution at thermal equilibrium and quasi electron distribution at higher electron energy respectively,$\tau_{b}$ and $\tau_{spon}$ are intra-band relaxation and spontaneous rate respectively.\\
Considering the entire spatial volume of the active region, we can write the rate of change of electron density and photon density in terms of the following variables, as seen in eq.\eqref{eq:rateeqn2}: Where~$n$ is carrier concentration(same as electron distribution~$N_{cb}$),~$I$ is the injected current,~$q$ is the charge,~$V$ is the volume of the active region,~$d$ is the thickness of the active region,~$\tau_{n}$ is the carrier lifetime,~$\tau_{p}$ is the photon lifetime in the cavity,~$\tau_{mode}$ is the radiative lifetime,~$n_{g}$ is the carrier concentration when the active region becomes transparent,~$\epsilon$ is the gain compression factor,~$a$ is the tangential coefficient,~$N_{s}$ is the photon concentration,~$\Gamma$ the mode confinement factor.

Our analysis evaluates:(I) Small signal excitation about steady-state value such that laser injection current follows step response near the threshold. Here, excitation is maintained above the threshold for a specific duration before turning off. Once the excitation is turned off, a steady-state value reached is below the threshold. This use-case emulates the behavior of the laser diode near its threshold region. (II) Another scenario is the large-signal excitation of the laser diode where the laser is pre-biased near the threshold, and then excited by a short-duration, large-signal excitation. This also models the behavior of the laser diode in [1] if the pre-bias current set to zero.
From eqn.\eqref{eq:rateeqn1}, as discussed in [8],  we can write the rate of change of electron distribution and photon output density for a single resonating mode as:

\begin{eqnarray}\label{eq:rateeqn2}
\dfrac{dn}{dt}=-\dfrac{\Gamma a(n-n_{g})}{1+\epsilon N_{0}}N_{s}-\dfrac{n}{\tau_{n}}+\dfrac{I}{qV}\\
\dfrac{dN_{s}}{dt}=\dfrac{\Gamma a(n-n_{g})}{1+\epsilon N_{0}}N_{s}+\dfrac{\Gamma \beta n}{\tau_{mode}}-\dfrac{N_{s}}{\tau_{p}}
\end{eqnarray}
where,
\begin{equation}\label{eq:rate}
\dfrac{1}{\tau_{n}}=\dfrac{1}{\tau_{nr}}+\dfrac{1}{\tau_{mode}}
\end{equation}\\
with ~$\tau_{nr}$ being non-radiative rate; in Eqn. \ref{eq:rateeqn2},~$\Gamma a $ is the gain coefficient,~$n$ is the electron distribution above the threshold where stimulated emission dominates and remains constant around the steady-state value.
For small-signal excitation, linearizing the change in electron distribution and output emission about its steady-state value, we can write:
\begin{eqnarray}\label{eq:diff1}
 \dfrac{d^{2}\Delta n}{dt^{2}}+[\dfrac{1}{\tau_{n}}+\dfrac{\Gamma aN_{s}}{1+\epsilon(N_{s}+\Delta N_{s})}]\dfrac{d\Delta n}{dt}\nonumber\\
 +\dfrac{\Gamma a N_{s}}{1+\epsilon(N_{s}+\Delta N_{s})}\dfrac{1}{\tau_{p}}\Delta n =0
\end{eqnarray}
Where $\Delta n$ is the small change in electron density in the conduction band and $\Delta N_{s}$ is the corresponding change in photon density. Since $\Delta N_{s}$ is smaller than $N_{s}$, it can be ignored. On solving the second-order differential equation eq.\eqref{eq:rate}, $\Delta n$ can then be written as:
\begin{equation}\label{eq:deltan}
    \Delta n=\Delta n_{0} e^{-{[\dfrac{1}{2}(\dfrac{1}{\tau_{n}}+\dfrac{\Gamma a N_{s}}{1+\epsilon N_{s}})-i\omega}]t}
\end{equation}
In eqn.\eqref{eq:deltan}, for~$t>0$, the secondary peak will attenuate with the increase in its real part. It is this behavior we tune in order to suppress secondary peaks. Here,
\begin{widetext}
\begin{equation}\label{eq:freq1}
  \omega=\sqrt{\dfrac{\Gamma a N_{s}}{1+\epsilon N_{s}}\dfrac{1}{\tau_{p}}-\dfrac{1}{\tau^{2}_{n}}-\dfrac{1}{4}[\dfrac{\Gamma a N_{s}}{1+\epsilon N_{s}}]^{2}-\dfrac{\Gamma a N_{s}}{2 \tau_{n}(1+\epsilon N_{s})}}
\end{equation}
\end{widetext}
if we represent the photon confinement within the cavity modes as $N_{p}$, we can then simplify~$\omega$ as:
\begin{equation}\label{eq:freq2}
  \omega=\sqrt{\dfrac{N_{p}}{\tau_{p}}-\dfrac{1}{\tau^{2}_{n}}-\dfrac{N_{p}^{2}}{4}-\dfrac{N_{p}}{2 \tau_{n}}}
\end{equation}
For sub-threshold region:
\begin{equation}\label{eq:Nsubthreshold}
    N_{s}=\dfrac{\tau_{n}\tau_{p}\Gamma \beta I}{\tau_{mode} qV}
\end{equation}
above the threshold:
\begin{equation}\label{eq:Nabovethreshold}
    N_{s}=\dfrac{\tau_{p}}{qd}[\dfrac{Id}{V}-\dfrac{n_{sp}qd}{\tau_{n}}]
\end{equation}
From eqn. \eqref{eq:deltan}-\eqref{eq:Nabovethreshold}, secondary fluctuation in electron density will reduce with increase in injection bias current at $t>0$.
Similar relation holds true for photon density and can be expressed as:
\begin{eqnarray}\label{eq:diff2}
 \dfrac{d^{2}\Delta N_{s}}{dt^{2}}+[\dfrac{1}{\tau_{n}}+\dfrac{\Gamma aN_{s}}{1+\epsilon(N_{s}+\Delta N_{s})}]\dfrac{d\Delta N_{s}}{dt}\nonumber\\
 +\dfrac{\Gamma a N_{s}}{1+\epsilon(N_{s}+\Delta N_{s})}\dfrac{1}{\tau_{p}}\Delta N_{s} =0
\end{eqnarray}
on neglecting the $\Delta N_{s}$
\begin{equation}\label{eq:deltaN}
    \Delta N_{s}=\Delta N_{0}e^{-[{\dfrac{1}{2}(\dfrac{1}{\tau_{n}}+\dfrac{\Gamma a N_{s}}{1+\epsilon N_{s}})-i\omega}]t}
\end{equation}
When excitation current below the threshold is injected into the laser diode, carrier concentration increases from certain initial value $n_{i}$ to a final value $n_{f}$. This change in carrier concentration has some associated rise time, $t_{r}$. When increase in this carrier concentration is such that rate of spontaneous emission is more than the cavity loss in the active medium, based on eqn.(23) the photon concentration gradually increases. Above the threshold region, electron concentration can be approximated to a constant value $n_{constant}$. This rise time for the electron concentration where photon emission begins, can be expressed as:
\begin{equation}\label{eq:delay}
    t_{r}=\tau_{n} ln\dfrac{I-\dfrac{qVn_{i}}{\tau_{n}}}{I-\dfrac{qVn_{constant}}{\tau_{n}}}
\end{equation}
where $\tau_{n}$ represents carrier lifetime and eq. \eqref{eq:delay} can be interpreted as follow: if the injected current is very high w.r.t the threshold, this rise time or characteristic delay in output emission theoretically approaches zero. Moreover, any residual delay otherwise can be adjusted by a tunable laser driver[21][24].

To further characterize emission from the laser diode rigorously: How do we prove that the output of such laser diodes can be realized as a (weak) coherent source? What is its origin and dynamics? \\ \\
If the dipole moment~$d$ vector due to the injection current is written as:
\begin{equation}\label{eq:dipole}
    d(t)=e(r_{0}e^{i\omega t}+r^{*}_{0}e^{-i\omega t})
\end{equation}
where~$r_{0}$ is the position vector associated with the electron and is expressed as a real quantity,
then interaction hamiltonian from eqn.\eqref{eq:dipole}  can be written in terms of quantized electric field operator~$E_{t}$ as:
\begin{equation}\label{eq:iHamiltonian}
    H_{interaction}(t)=d(t).E_{t}
\end{equation}
If~$\kappa$ is a propagation vector,~$\epsilon$ as the polarization modes, and~$V$ as quantization of the volume; we can write the quantized transverse electric field as:
\begin{equation}\label{eq:electricfield}
    E_{t}=i\sum_{\kappa, \epsilon_{k}}\sqrt{\dfrac{\hbar\omega_{k}}{2\epsilon_{0} V}}(a_{\kappa,\epsilon_{\kappa}} \epsilon_{\kappa} e^{i\kappa \cdot R}-a_{\kappa, \epsilon_{\kappa}}^{\dagger}\epsilon_{\kappa}e^{-i \kappa \cdot R})
\end{equation}
with the Hamiltonian for the charge particle and free field is written as:
\begin{equation}\label{eq:nonHamiltonian}
    H_{0} = -\dfrac{\hbar^{2}}{2m_{e}}\nabla^{2}_{Re} -\dfrac{\hbar^{2}}{2m_{h}}\nabla^{2}_{Rh}-\dfrac{e^{2}}{\epsilon|r_{e}-r_{h}|}+\sum_{\kappa}\hbar\omega_{\kappa}(a^{\dagger}_{\kappa}a_{\kappa}+\dfrac{1}{2})
\end{equation}
is also the eigenvalue equation:
\begin{equation}\label{eq:eignevalue}
    H_{0}|n\rangle = E_{n} |n\rangle
\end{equation}
so the overall Hamiltonian can be written as:
\begin{equation}\label{eq:quantizedelectric}
\begin{aligned}
  H_{total} &=H_{0}-d(t)\cdot i\sum_{\kappa, \epsilon_{k}}\sqrt{\dfrac{\hbar\omega_{k}}{2\epsilon_{0} V}}(a_{\kappa,\epsilon_{\kappa}} \epsilon_{\kappa} e^{i\kappa \cdot R}  \\
    &\quad -a_{\kappa, \epsilon_{\kappa}}^{\dagger}\epsilon_{\kappa}e^{-i \kappa \cdot R})
\end{aligned}
\end{equation}
 In the Dirac interpretation, Hamiltonian can then be written as:
 \begin{equation}\label{eq:interactionpicture}
     H_i(t)=e^{iH_{0}t/\hbar} H(t) e^{-iH_{0}t/\hbar}
 \end{equation}
\begin{equation}\label{eq:interactionhamiltonian}
\begin{aligned}
  H_{i}(t)&= -d(t)\cdot i\sum_{\kappa, \epsilon_{k}}\sqrt{\dfrac{\hbar\omega_{k}}{2\epsilon_{0} V}}(e^{iH_{0}t/\hbar}a_{\kappa,\epsilon_{\kappa}}e^{-iH_{0}t/\hbar} \epsilon_{\kappa} e^{i\kappa \cdot R}\\
         &\quad  -e^{iH_{0}t/\hbar}a_{\kappa, \epsilon_{\kappa}}^{\dagger}e^{-iH_{0}t/\hbar}\epsilon_{\kappa}e^{-i \kappa \cdot R}) 
\end{aligned}
\end{equation}
In the interaction picture, the time evolution of the quantum state is obtained from the time-dependent Schrodinger equation:
\begin{equation}\label{eq:Schrodinger}
    i\hbar\dfrac{d}{dt}|\psi(t)\rangle=H(t)|\psi(0)\rangle
\end{equation}
considering vacuum state as the initial state, the solution to eq.\eqref{eq:Schrodinger} can be represented as:
\begin{equation}\label{eq:Unitary}
    |\psi(t)\rangle=U(t,t_{0})|0\rangle
\end{equation}
Due to injection current profile, Hamiltonian at time~$t_{1}$ and~$t_{2}$ is such that~$[H(t_{1}),H(t_{2})]\neq0$
then, U(t,$t_{0}$) can be expressed in terms of Dyson series as:
\begin{equation}\label{eq:Dyson}
\begin{aligned}
    U(t,t_{0})&=I+\sum_{n=0}^{\infty}[-\dfrac{i}{\hbar}]^{n}\int_{t_{0}}^{t}dt_{1}\int_{t_{0}}^{t_{1}}dt_{2}\cdot\cdot\cdot\cdot\\
              &\quad\int_{t_{0}}^{t_{n-1}}dt_{n}H(t_{1})H(t_{2})\cdot\cdot\cdot\cdot~H(t_{n})
\end{aligned}
\end{equation}
From Eqn.\eqref{eq:Dyson}, the limit of the integration determines the integration variable in a later instance for $t_{n}$ and so on. In order to avoid this, we can re-write it in terms of time product of operators as:
\begin{equation}\label{eq:timeorder}
\begin{aligned}
    U(t,t_{0})&=I+\sum_{n=0}^{\infty}\dfrac{1}{n!}[-\dfrac{i}{\hbar}]^{n}\int_{t_{0}}^{t}dt_{1}\int_{t_{0}}^{t}dt_{2}\cdot\cdot\cdot\cdot\\
              &\quad\int_{t_{0}}^{t}dt_{n}T[H(t_{1})H(t_{2})\cdot\cdot\cdot\cdot~H(t_{n})]
\end{aligned}
\end{equation}
or,
\begin{equation}\label{eq:timeorderexponential}
    U(t,t_{0})=Te^{-\dfrac{i}{\hbar}\int_{t_{0}}^{t}H(t)dt}
\end{equation}
where~$T$ is the time ordered product operator.
\begin{equation}\label{eq:quantumstatetimeorder}
    |\psi_i(t)\rangle = Te^{-\dfrac{i}{\hbar}\int_{t_{0}}^{t}H(t)dt}|0\rangle
\end{equation}
and exponential term can be expressed following the approach discussed in[18]:
\begin{equation}\label{eq:creationoperator}
    -\dfrac{i}{\hbar}\int_{0}^{t}H_{i}(t)dt=\sum_{\kappa,\epsilon_{k}}[-\alpha^{*}_{k}(t)a_{k}+\alpha_{k}(t)a^{\dagger}_{\kappa}]
\end{equation}
where
\begin{equation}\label{eq:conjugate}
    \alpha^{*}_{k}(t)=\dfrac{1}{\hbar}\sqrt{\dfrac{\hbar\omega_{k}}{2\epsilon_{0} V}}\int_{0}^{t}dt  d(t)(e^{iH_{0}t/\hbar}a_{\kappa,\epsilon_{\kappa}}e^{-iH_{0}t/\hbar} \epsilon_{\kappa} e^{i\kappa.R})
\end{equation}\\
from eqn. \eqref{eq:quantumstatetimeorder}-\eqref{eq:conjugate}:\\
\begin{equation}\label{eq:coherentstate}
 |\psi_{i}(t)\rangle = Te^{\sum_{\kappa, \epsilon_{k}}[-\alpha_{k}^{*}(t)a_{k}+\alpha_{k}(t)a_{\kappa}^{\dagger}]}|0\rangle
 \end{equation}
 or
 \begin{equation}\label{eq:displacement}
  |\psi_{i}(t)\rangle =T\prod_{\kappa}e^{[-\alpha_{k}^{*}(t)a_{k}+\alpha_{k}(t)a_{\kappa}^{\dagger}]}|0\rangle
 \end{equation}
in terms of the displacement operator:
\begin{equation}\label{eq:displacementoperator}
    |\psi_{i}(t)\rangle=\prod_{\kappa}D_{\kappa}|0\rangle
\end{equation}
From Eqn. \eqref{eq:displacementoperator}, in a laser diode, displacement operation on a vacuum state due to current injection results in multi-mode coherent state emission. To understand dynamics of evolution of the vacuum state, if we define the scattering matrix as:
\begin{equation}\label{eq:smatrix}
    S(t',t)=U(t,0)U^{\dagger}(t',0)
\end{equation}
we can write S(t',t) as:
\begin{equation}\label{eq:smatrixasdysonseries}
\begin{aligned}
    S(t',t)&=I+\sum_{n=0}^{\infty}\dfrac{1}{n!}[-\dfrac{i}{\hbar}]^{n}\int_{t'}^{t}dt_{1}\int_{t}'^{t}dt_{2}\cdot\cdot\cdot\cdot\\
           &\quad\int_{t'}^{t}dt_{n}T[H(t_{1})H(t_{2})\cdot\cdot\cdot\cdot~H(t_{n})]
\end{aligned}
\end{equation}
We can then define two particle Zero-temperature Green's function(or Causal Green's function):
\begin{equation}\label{eq:greenfunction}
\begin{aligned}
    G(x_{1},t_{1}&;x_{2},t_{2};x_{1}^{`},t_{1}^{`};x_{2}^{`},t_{2}^{`})=-i\langle GS|T[a_{x_{2}^{'}}(t_{2}^{'})\\
    &a_{x_{1}^{'}}(t_{1}^{'})a_{x_{1}}^{\dagger}(t_{1})a_{x_{2}}^{\dagger}(t_{2})]|GS\rangle
      \end{aligned}
\end{equation}
Since ground state of non-interacting Hamiltonian is calculated from Schrodinger equation, it is an eigen-state of non-interacting Hamiltonian. So, $\langle GS|$, the eigen state of overall Hamiltonian, comprising of non-interacting and interaction Hamiltonian, is then estimated using Gell-Mann and Low equation[20]:
\begin{equation}\label{eq:Gellmann}
    |GS\rangle=\hat S(0,-\infty)|0\rangle
\end{equation}
For electron-hole pair(EHP), eqn.\eqref{eq:greenfunction} can be written in terms of creation and annihilation operator as:
\begin{equation}\label{eq:greenfunctiontimeordered}
\begin{aligned}
    G(x^{'},t^{'};x,t)=-i\langle GS|T[a_{x^{'}}(t^{'})
    a_{x^{'}}(t^{'})a_{x}^{\dagger}(t)a_{x}^{\dagger}(t)]|GS\rangle
      \end{aligned}
\end{equation}

Physical interpretation of Green's function[19] for EHP can be visualized from eqn.\eqref{eq:greenfunctiontimeordered}. EHP is created at time ~$t(t^{'}> t)$. EHP then propagates through the system as the resulting state is not the overall Hamiltonian eigenstate. At~$t'$, EHP is removed from the system by annihilation operators.
Writing operators in an interaction picture, we can then write Green's function in terms of S-matrix, with electron and hole momentum as~$|k|$ as:
\begin{equation}\label{eq:greenfunctionsmatrix}
\begin{aligned}
   &G(x^{'},t^{'};x,t)=-i\langle 0|T[\hat S(0,\infty)\hat S(\infty,t)a_{-k}(t^{'})a_{k}(t^{'})\hat S(t,t')\\
   &a_{-k}^{\dagger}(t)a_{k}^{\dagger}(t)]\hat S(t',0)\hat S(0,-\infty)|0\rangle/(\langle 0|\hat S(\infty,-\infty)|0\rangle)
      \end{aligned}
\end{equation}
Feynman diagram for spontaneous emission can then be represented from eqn.\eqref{eq:greenfunctionsmatrix} as:
\begin{figure}[ht!]
\includegraphics[width=6cm]{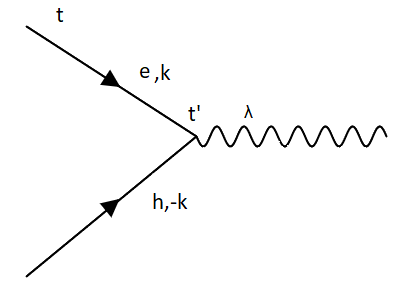}
\caption{\emph{Feynman diagram for spontaneous emission from EHP recombination. EHP with momentum~$|k|$ annihilates at time~$t'$ to emit photon of wavelength $\lambda$.}}
\label{fig:feynmandiagramspn}
\end{figure}
An EHP with momentum~$|k|$ annihilates at time~$t'$ to emit a photon of wavelenght~$\lambda$.
When the rate of emission builds up, photon density increases before a steady value of carrier density and photon density is reached. Then an increase in photon density results in the reduction of EHP due to recombination. Further, the spatial coherence of the emission is ensured by selecting the cavity such that the emission wavelength is near the spectral peak, resulting in the narrow spectrum of the laser emission.
\section{Numerical Simulation}

\begin{figure}[ht!]
\includegraphics[width=9.5cm]{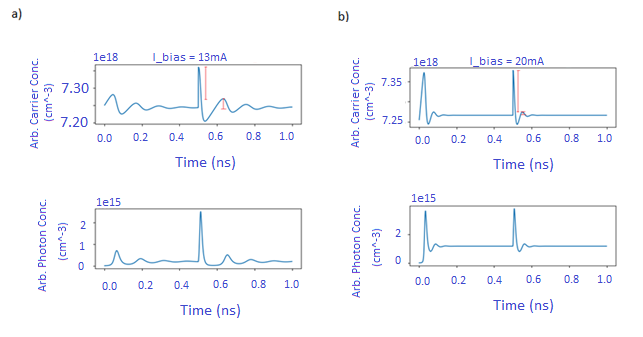}
\caption{\emph{Simulation of relaxation oscillation for different values of injection current. Fig.a represents change in carrier concentration and photon concentration when the laser diode is excited by a short duration pulse, with the pre-bias current of 13mA i.e. below the threshold. Fig.b represents a change in carrier concentration and photon concentration when the laser diode is excited by a short duration pulse, with the pre-bias current of 20mA i.e. above the threshold. As seen from Fig.a and Fig.b, the amplitude of the secondary oscillation peak decreases when the pre-bias current is increased. Also, the difference between primary peak and secondary peak increases.}}
\label{fig:impulseexcitation}
\end{figure}

We apply a large step excitation to the laser diode under study. This condition was simulated [25] by solving the rate equation using Euler’s iterative method for parameters defined as:

\begin{figure}[ht!]
\includegraphics[width=9cm]{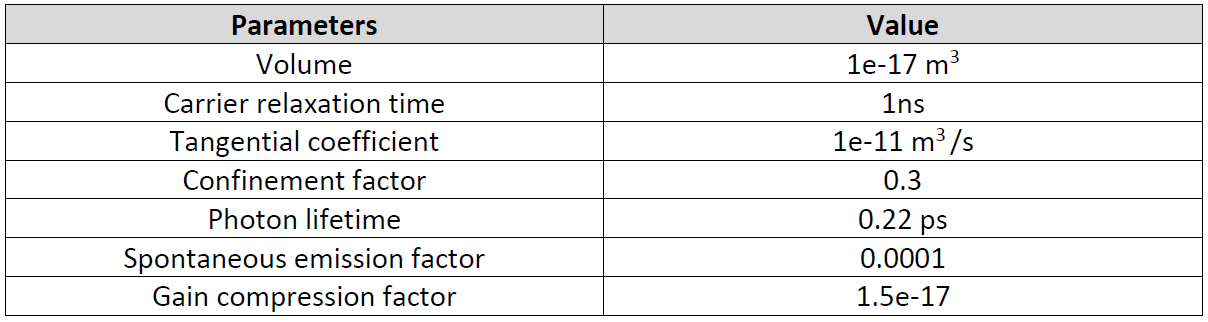}
\caption{\emph{Parameter defined for the laser diode simulation}}
\label{fig:parameters}
\end{figure}
\begin{figure}[ht!]
\includegraphics[width=8cm]{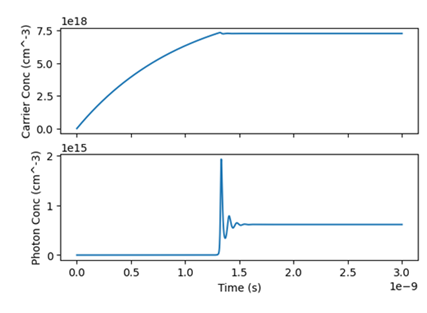}
\caption{\emph{Simulation of large signal step function: If the step function is turned off at ~1.5ns, the responses will resemble a pulse excitation with two peaks.}}
\label{fig:relaxationoscillationsimulation}
\end{figure}
Fig. \ref{fig:relaxationoscillationsimulation} represents the response of the carrier and photon concentration when the laser diode is excited by the large-signal step function. It can be observed that photon concentration follows the carrier concentration graph with some delay. Had this excitation been turned off at 1.5ns, it would have resembled the excitation profile described in [1]. In order for the decoy-state described in [1] to be indistinguishable from the signal state, the secondary peak of the signal state must cease to exist; otherwise, the decoy-state will have one peak, while the signal will have two peaks. Following the trend predicted by eq.\eqref{eq:deltan}and eq.\eqref{eq:deltaN} for small signal excitation and Fig. \ref{fig:relaxationoscillationsimulation} for large step excitation, Fig. \ref{fig:impulseexcitation} represents the solution to the rate equation under a large-signal excitation profile with a short duration pulse and a non-zero bias current.\\
We have considered two simulation models: In the first model the constant pre-bias current defined as $< I_{threshold}$, Fig. \ref{fig:impulseexcitation}.a. In the second model, pre-bias current defined as $> I_{threshold}$, Fig. \ref{fig:impulseexcitation}.b. No other simulation parameters were varied. Perturbation of duration 2ps was then superimposed on the pre-bias current, and relaxation oscillation due to the perturbation was characterized. It was ensured that the perturbation is applied at $t>>0$ so that the response due to perturbation is not affected by initial startup transients. In Fig. \ref{fig:impulseexcitation} the amplitude of secondary oscillation decreases with an increase in injection current, and the difference between primary amplitude to amplitude of secondary oscillation increases with an increase in injection current. This result trends with the estimation made in section 2 for the small signal consideration (Initial transients seen in the waveform are due to switching of pre-bias current). Consequently,
to suppress the peak of the secondary oscillation, the pre-bias current must be increased. Any residual peak can be filtered by the driver circuit. 
\section{Experiment}
\begin{figure}[ht!]
\includegraphics[width=9cm]{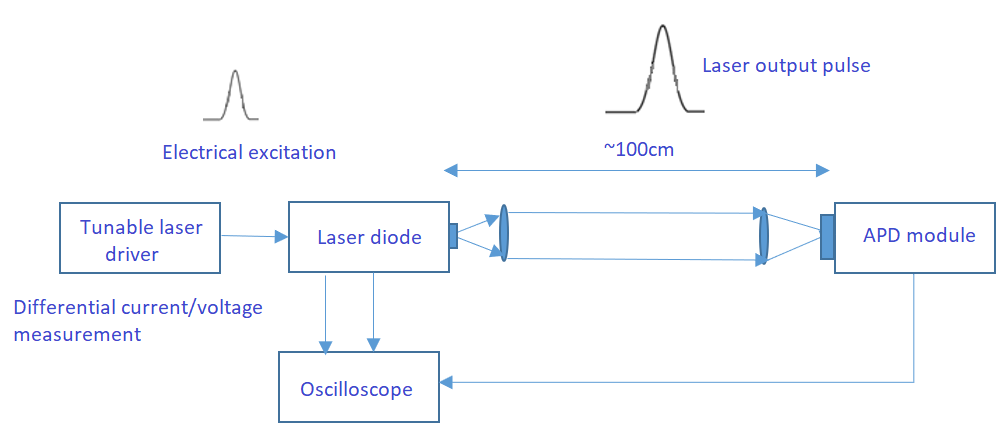}
\caption{\emph{Experimental setup for the characterization of laser diode: Emission from a tuned laser diode is detected using an APD biased in linear region. The differential current into the laser diode and response from the APD are measured by an oscilloscope. Threshold current of the laser diode is ~18.4mA.}}
\label{fig:experimentalsetup}
\end{figure}
The measurement setup as shown in Fig.\ref{fig:experimentalsetup} comprises of a transmitter: we developed a tunable laser driver[21][24] capable of generating dc bias superposed on modulated sub-nanosecond injection current into a HL6748MG laser diode at room temperature. Our intended application aims to have a laser diode as cheap as possible and have a minimal form factor. Measurement has been done without using a constant optical attenuator because weak photon pulses in a single-photon regime will follow the same characteristics as the un-attenuated pulse.The polarization basis of the photon is selected using a polarizer and half-wave plate. For detection, we developed a SAP500 APD-based detector capable of measuring photons in linear mode and Geiger mode.\\
Injection current through the laser diode was measured using differential probe D420-A-PB with 4GHz DX20 tip connected to 4GHz, Lecroy 640Zi Waverunner. This interface has a rise time of 122.5ps. The output of APD was interfaced to the same oscilloscope with active probe ZS2500, 2GHz bandwidth; the rise time of the interface was ~175ps. Additionally, considering the 500ps rise/fall time of the APD, sharp peaks in the ideal waveform of the period ~300ps will still be observed, albeit not as sharp as reported in [1]. 
\section{Results}
\begin{figure}[ht!]
\includegraphics[width=9cm]{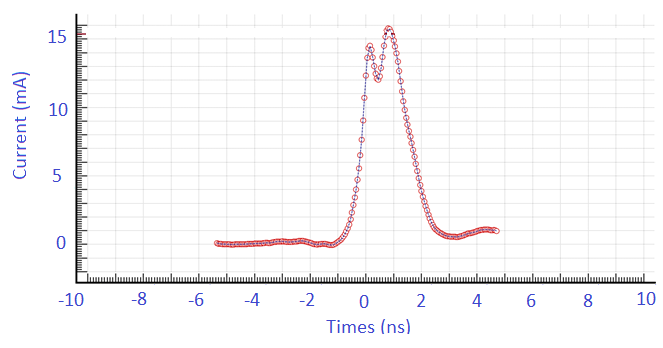}
\caption{\emph{Differential profile across laser diode terminal. The graph has been averaged over 1500 samples}}
\label{fig:nobiasrelaxationoscillation}
\end{figure}

Fig. \ref{fig:nobiasrelaxationoscillation} represents a differential current profile measured across terminals of the laser diode when excited by a peak current pulse of near-threshold amplitude. A secondary peak due to relaxation oscillation at 800ps can be observed.
Bandwidth limitation in the setup will affect the absolute accuracy of rise time and periods of the peaks due to the comparatively slow rise time of the probe cable and the detector. In Fig. \ref{fig:nobiasrelaxationoscillation}, we are expecting relaxation oscillation then why the secondary peak is higher than the primary peak? Our measuring probe rise time (122.5ps) is more than the pulse's actual rise/fall time which add the positive offset in the measurement.
In order to produce an indistinguishable signal state and decoy-state, this profile must be indistinguishable, i.e., a secondary peak in this waveform must vanish.

\begin{figure}[ht!]
\includegraphics[width=9cm]{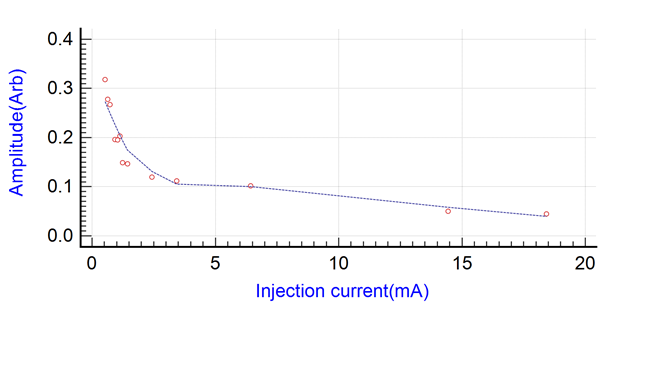}
\caption{\emph{Variation of oscillation peak as a function of the injected dc bias current (without filters). As observed, the secondary oscillation peak is still non-zero. This secondary peak must be suppressed to zero and should approach to zero or below.}}
\label{fig:measurementwtfiltersoscillationpeak}
\end{figure}

\begin{figure}[ht!]
\includegraphics[width=9cm]{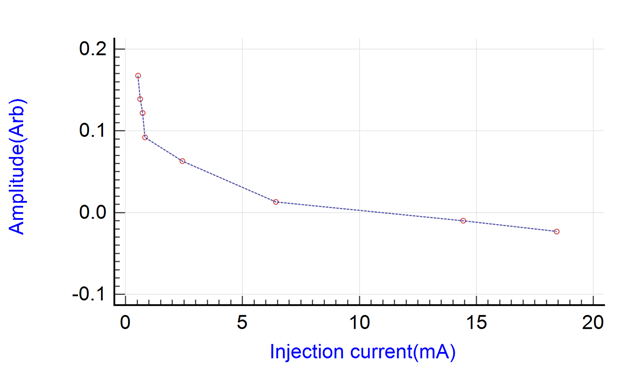}
\caption{\emph{Variation of the oscillation peak as a function of injected dc bias current (with filters). Non-zero peak is suppressed at approximately 11mA.}}
\label{fig:measurementwfiltersoscillationpeak}
\end{figure}

\begin{figure}[ht!]
\includegraphics[width=9cm]{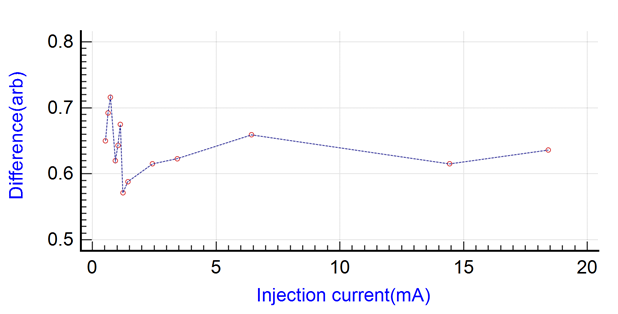}
\caption{\emph{Variation of difference in the peak of signal and secondary oscillation (without filters). As the injection current increases and approaches the threshold level, difference between primary and secondary oscillation peak increases.}}
\label{fig:measurementwtfilterspeakdifference}
\end{figure}

\begin{figure}[ht!]
\includegraphics[width=9cm]{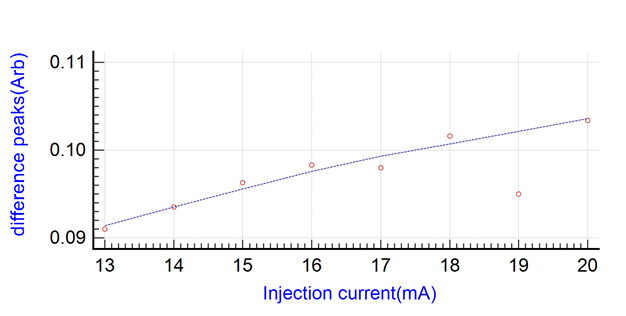}
\caption{\emph{Simulated for variation of difference in the peak of signal and secondary oscillation. As compared with Fig.\ref{fig:measurementwtfilterspeakdifference}, trend observed in the measurement result is corroborated in simulation. Difference between the amplitude values for the measurement and the simulation can be attributed to deviation in the simulation model from the actual device value.}}
\label{fig:simulationpeakdifference}
\end{figure}

Fig. \ref{fig:measurementwtfiltersoscillationpeak} represents the variation of the peak of secondary oscillation as a function of injected current (no additional electronic filter was implemented). As observed, the magnitude of the injected bias current and amplitude of the secondary peak are inversely related. For the injected bias current above the threshold, the amplitude of the secondary peak is non-zero. On implementing an electronic filter in the laser driver, the secondary peak was further reduced and approached zero for the injected current of 11mA (Fig.\ref{fig:measurementwfiltersoscillationpeak}).
Difference between the amplitude of the primary and secondary peak increases with the biased current near the threshold value (Fig.\ref{fig:measurementwtfilterspeakdifference}) because of reduction in the amplitude of the secondary peak and increase in the magnitude of the primary peak.
Fig.\ref{fig:simulationpeakdifference} represents numerical simulation result for estimating variation in difference between primary and secondary peaks which increases with the injection current.
\begin{figure}[ht!]
\includegraphics[width=9cm]{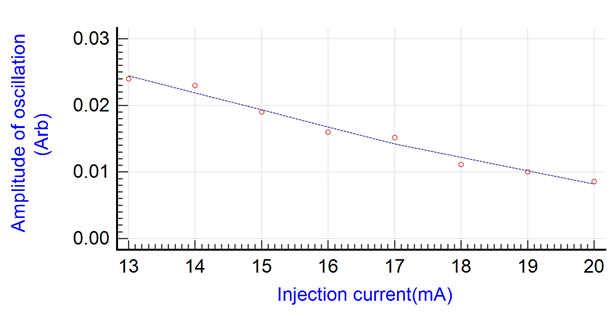}
\caption{\emph{Variation of peak of secondary oscillation (simulation result). Trend observed in measurement, as seen in Fig. \ref{fig:measurementwfiltersoscillationpeak} and \ref{fig:measurementwtfiltersoscillationpeak} is corroborated by simulation.}}
\label{fig:simulationpeak}
\end{figure}
Further, Fig. \ref{fig:simulationpeak} represents numerical simulation result of variation in the amplitude of the secondary peak which reduces with increase in the injection current into the laser diode.
\section{Discussions}
\begin{figure}[ht!]
\includegraphics[width=9cm]{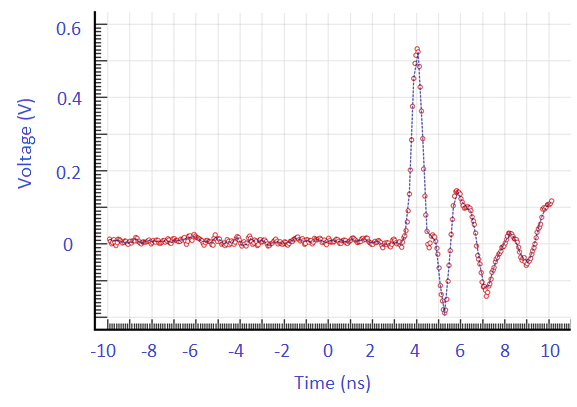}
\caption{\emph{APD output for the laser diode biased near to the threshold at 19.3mA bias current. Multiple oscillations are due to large inductance in the receiver.Voltage drop across APD series resistor of 500K was 10V which corresponds to 200uA of APD current. With sub-nanosecond peak pulse emission from the laser diode, power ~120mW. Considering overall coupling efficiency of 2.5$\%$, with 50 ohm APD load resistor and responsivity of 14.8A/W for 672nm wavelength, expected output voltage on the oscilloscope is around 555mV.}}
\label{fig:signalAPD}
\end{figure}

Voltage drop across the APD series resistor of 500K was 10V, corresponding to the maximum allowed of the APD current of 200uA. With sub-nanosecond peak pulse emission from the laser diode, with power ~120mW, overall coupling efficiency of 2.5$\%$, APD load resistor in the range of around 50 ohm and responsivity of 14.8A/W for 672nm wavelength, expected output voltage on the oscilloscope considering attenuation due to probes, is around 555mV. So, the plot in Fig.\ref{fig:signalAPD} looks reasonable.
\begin{figure}[ht!]
\includegraphics[width=9cm]{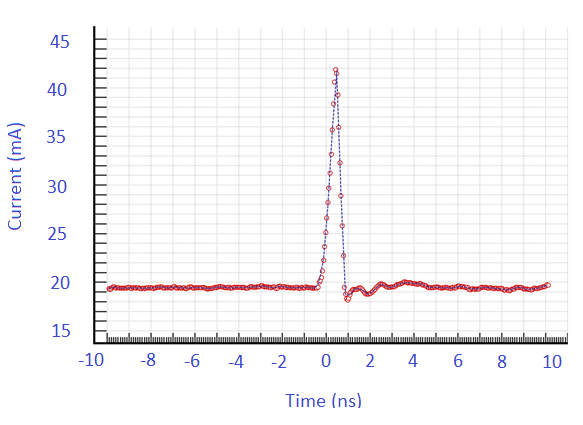}
\caption{\emph{Current profile for the signal state measured across the laser diode. Pre-bias current is maintained above the laser threshold level. Peak current injected into the laser diode is less than the absolute maximum rating of 45mA.}}
\label{fig:signalexcitation}
\end{figure}

\begin{figure}[ht!]
\includegraphics[width=9cm]{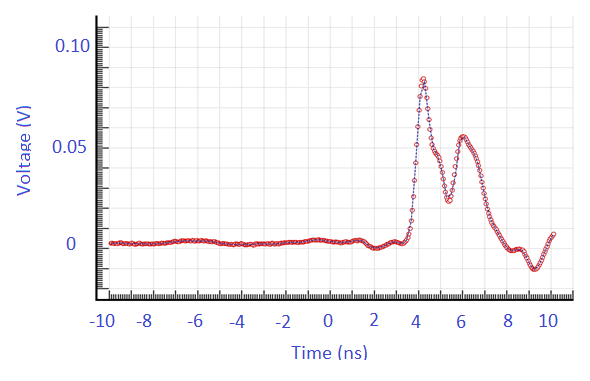}
\caption{\emph{APD output for the laser diode biased near the threshold (at 15mA bias current). Multiple oscillations are due to large inductance and appearance of double peak above the threshold is due to delay introduced by the oscilloscope probe. Profile with the pre-bias current of 19.3mA (figure 6) and profile for the pre-bias current of 15mA are the same}}
\label{fig:decoyAPD}
\end{figure}

Fig. \ref{fig:decoyAPD} represents APD output response for the laser diode pre-biased near the threshold current level. Comparing it with Fig. \ref{fig:signalAPD}, light pulse profile looks similar, hence indistinguishable in shape. Side bumps observed in Fig.\ref{fig:signalAPD} and \ref{fig:decoyAPD} are due to parasitic on the transmitter and receiver PCB. Limited bandwidth of the APD and probes results in some positive offset instead of expected zero line at 5ns.
\begin{figure}[ht!]
\includegraphics[width=9cm]{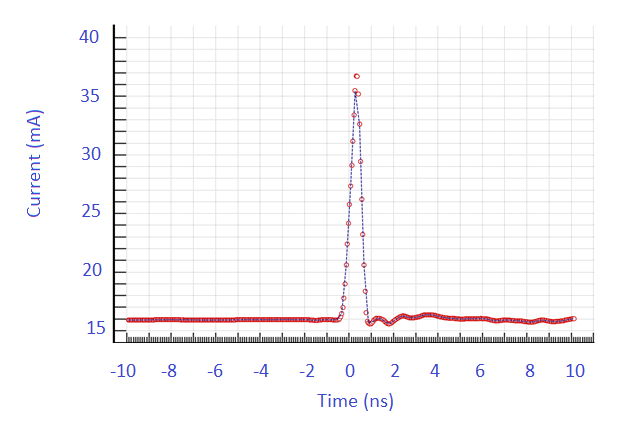}
\caption{\emph{Current injection profile for the decoy state measured across the laser diode. Pre-bias level is maintained near to the threshold level, with peak current reaching upto 37mA.}}
\label{fig:decoyexcitation}
\end{figure}

Another factor that can distinguish these pulses is the delay between the signal and the decoy-state. As observed from eqn. \eqref{eq:delay}, delay between the laser excitation and the laser output depends on the current injection. From Fig.\ref{fig:signalAPD} and Fig. \ref{fig:decoyAPD}, relative delay between two output pulses is around 400ps. This can be attributed to the difference in injection current which governs the delay between carrier density and photon emission. Further, from Fig.\ref{fig:signalexcitation} and Fig.\ref{fig:signalAPD} the absolute delay between the laser excitation pulse and output of the APD is in the range of ns. This can be attributed to the distance of the APD from the laser (3.3ns), APD response time, and the rise time of the probe. Further, in the implementation, decoy state can be achieved by varying the pre-bias current level or the magnitude of the perturbation pulse[24].
\begin{figure}
    \centering
    \includegraphics[width=9cm]{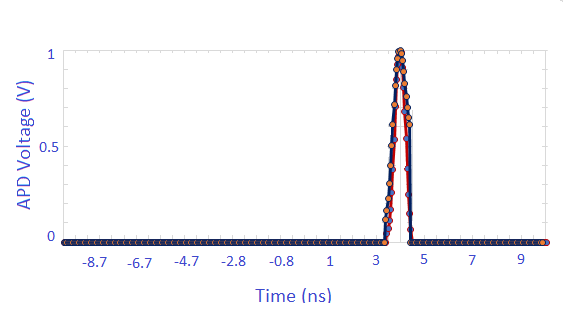}
    \caption{\emph{Comparison of the signal state and the decoy state. Graphs have been amplitude normalized. Relative delay between the response has been neglected so that peaks of both the responses are temporally aligned for better comparison. Any PCB relevant parasitic oscillations have been eliminated at the zero reference level}}
    \label{fig:compare}
\end{figure}

\begin{figure}
    \centering
    \includegraphics[width=9cm]{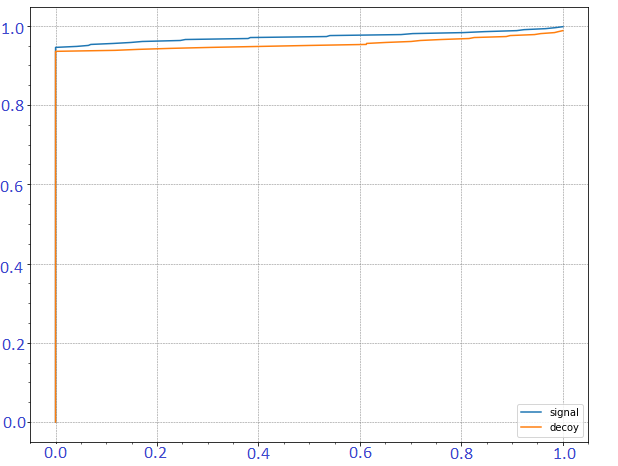}
    \caption{\emph{Cumulative distribution function for the signal and the decoy state. Based on pvalue=0.999664050220288 and statistic=0.024875621890547265, it can be concluded that both the signal and the decoy state data belong to same the distribution.}}
    \label{fig:Kolmogorov}
\end{figure}

Further, we statistically evaluated the distributions shown in Fig.\ref{fig:compare}, using the Kolmogorov-Smirnov test the estimated p-value is 0.999664 greater than 0.05. So we can accept the hypothesis that both of these responses came from the same distribution, i.e., both are indistinguishable. Additionally, Fig.\ref{fig:Kolmogorov} represents the cumulative distribution function for the signal and the decoy-state.

In this paper, we have considered indistinguishability only in the temporal domain. Distinguishability in the spectral domain during a transient process can occur due to pump current excitation or when an external optical modulator is used for optical pulse generation[7]. It has been observed that the shift depends on the magnitude of current change, laser pre-bias condition, and oscillation damping rate. It will of great interest to understand how the pre-bias current in our approach reduces drift in the wavelength of the signal and the decoy state.
\section{Conclusion}

We analyzed the cause of distinguishability between the signal and the decoy-state in a pump modulated laser diode. Theoretical derivation about the vacuum state evolving into stimulated emission characterizes coherent state emission from the laser diode. We have further demonstrated indistinguishability of the signal and the decoy-state by suppressing relaxation side peaks. The proposed relaxation oscillation suppression technique will be of interest in implementing two-photon interference [23] and reducing transient chirps for indistinguishability in the frequency domain.
\section{Acknowledgement}
This study is funded by Robert Bosch LLC. The author would like to thank Thomas Strohm, Corporate Research Group, Robert Bosch GmbH for helpful discussions.

\section{References}
\begin{enumerate}
    \item Huang, A., Sun, S.-H., Liu, Z., and Makarov, V., Physical Review A, vol. 98, no. 1,(2018) doi:10.1103/PhysRevA.98.012330
    \item Sebastian Nauerth et al 2009 New J. Phys. 11 065001
    \item Goldwasser, S., Tauman Kalai, Y. (2016) Springer, Berlin, Heidelberg
    \item Nicolas Sangouard \& Hugo Zbinden (2012), Journal of Modern Optics, 59:17, 1458-1464
    \item Hoi-Kwong Lo and John Preskill. 2007 Quantum Info. Comput. 7, 5 (July 2007), 431–458.
    \item D. Gottesman, H.-K. Lo, N. L¨utkenhaus, \& J. Preskill, Quantum Info. and Comp., 4, No.5 (2004) 325-360
    \item R. Linke, IEEE Journal of Quantum Electronics, vol. 21, no. 6, pp. 593-597, June 1985, doi: 10.1109/JQE.1985.1072705
    \item P. Bhattacharya, Semiconductor optoelectronics devices, Prentice Hall 
    \item Dixon AR, Dynes JF, Lucamarini M, et al. Sci Rep. 2017;7(1):1978. Published 2017 May 16. doi:10.1038/s41598-017-01884-0
    \item Hua Lu, J. Opt. Soc. Am. B 36, B26-B30 (2019)
    \item Kiyoshi Tamaki, Marcos Curty and Marco Lucamarini, New J. Phys. 18 065008, (2016)
    \item X Wang:Phys. Rev. Lett. 94, 230503 (2005)
    \item M. Yamda, Theory of Semiconductor Lasers, Springer, 2014
    \item Richard J Hughes et al 2002 New J. Phys. 4 43
    \item Hoi-Kwong Lo, Xiongfeng Ma, and Kai Chen, Phys. Rev. Lett. 94, 230504 (2005)
    \item Maan, Pranshu “Resonant Fluorescence Spectroscopy in Low Dimensional Semiconductor Structures.” (2017)
    \item P. W. Shor, Proceedings 35th Annual Symposium on Foundations of Computer Science, 1994, pp. 124-134, doi: 10.1109/SFCS.1994.365700
    \item Principles of Laser spectroscopy and Quantum Optics: Berman and Malinovsky - Princeton university press
    \item Thomas Strohm PhD. Thesis, Nov 2004
    \item M.Gell-Mann and F.Low, Bound states in quantum field theory, Phys.Rev.84, 350(1951)
    \item Maan, P. (2022). U.S. Patent No. 11,233,579. Washington, DC: U.S. Patent and Trademark Office.
    \item Xiongfeng Ma, Bing Qi, Yi Zhao, and Hoi-Kwong Lo, Phys. Rev. A 72, 012326 (2005)\\
    \item R. Shakhovoy, V. Sharoglazova, A. Udaltsov, A. Duplinskiy, V. Kurochkin and Y. Kurochkin,IEEE Journal of Quantum Electronics, April 2021
    \item Maan, P. (2022). Results in Optics, 6, 100198
    \item Adapted from Niall Boohan, 2018: Program to simulate laser-rate equation in Python
    
\end{enumerate}
\end{document}